\documentclass[aps,pre,twocolumn,groupedaddress,superscriptaddress, showpacs]{revtex4-1}

\usepackage{graphicx}
\usepackage{amsmath}
\usepackage{amsfonts}
\usepackage{amssymb}

\usepackage{color}

\begin{document}

\title{Conductivity of Coniglio-Klein clusters}

\author{Nicolas Pos\'e}
  \email{posen@ifb.baug.ethz.ch}
  \affiliation{Computational Physics for Engineering Materials, IfB, ETH Zurich, Schafmattstrasse 6, CH-8093 Zurich, Switzerland}
  
\author{N. A. M. Ara\'ujo}
  \email{nuno@ethz.ch}
  \affiliation{Computational Physics for Engineering Materials, IfB, ETH Zurich, Schafmattstrasse 6, CH-8093 Zurich, Switzerland}
  
\author{H. J. Herrmann}
\email{hans@ifb.baug.ethz.ch}
  \affiliation{Computational Physics for Engineering Materials, IfB, ETH Zurich, Schafmattstrasse 6, CH-8093 Zurich, Switzerland}
  \affiliation{Departamento de F\'isica, Universidade Federal do
Cear\'a, 60451-970 Fortaleza, Cear\'a, Brazil}
 
\begin{abstract}
We performed numerical simulations of the $q$-state Potts model to compute the reduced conductivity exponent $t/ \nu$ for the critical Coniglio-Klein clusters in two dimensions, for values of $q$ in the range $[1;4]$. At criticality, at least for $q<4$, the conductivity scales as $C(L) \sim L^{-\frac{t}{\nu}}$, where $t$ and $\nu$ are, respectively, the conductivity and correlation length exponents. For $q=1$, $2$, $3$, and $4$, we followed two independent procedures to estimate $t / \nu$. First, we computed directly the conductivity at criticality and obtained $t / \nu$ from the size dependence. Second, using the relation between conductivity and transport properties, we obtained $t / \nu$ from the diffusion of a random walk on the backbone of the cluster. From both methods, we estimated $t / \nu$ to be $0.986 \pm 0.012$, $0.877 \pm 0.014$, $0.785 \pm 0.015$, and $0.658 \pm 0.030$, for $q=1$, $2$, $3$, and $4$, respectively. We also evaluated $t /\nu$ for non integer values of $q$ and propose the following conjecture $40gt/ \nu=72+20g-3g^2$ for the dependence of the reduced conductivity exponent on $q$, in the range $ 0 \leq q \leq 4$, where $g$ is the Coulomb gas coupling.
\end{abstract}


\maketitle

\section{Introduction}\label{section::Introduction}

The $q$-state Potts model was initially developed to study the onset of ferromagnetic order, but its range of applications is much wider and includes, for example, problems in materials science \cite{Anderson84, Holm01}, biology \cite{Graner92}, opinion dynamics \cite{Araujo10b}, image processing \cite{Bentrem10}, and quantum chromodynamics \cite{Alford01}. The general interest for statistical physics stems from its rich critical behavior and the fact that it generalizes several other models like, e.g., spanning trees ($q=0$), bond percolation ($q=1$), and Ising model ($q=2$) \cite{Wu82}. The critical properties of the order-disorder transition in the Potts model depend on the dimensionality and this transition might even be absent on complex geometries \cite{Andrade05, Andrade09b, Araujo10c, Dorogovtsev08}. In particular, in two dimensions, the nature of the transition is second order for $q\leq4$ and first order for $q>4$.

The theorem of Kasteleyn-Fortuin shows that, for all real values of $q>0$, the magnetic transition can be described by a purely geometrical model, where the partition function is a sum over bond percolation configurations weighted by a factor depending on the number of possible states and clusters \cite{Fortuin72}. This beautiful result has provided the necessary ingredients to develop advanced analytic and computational techniques to characterize the critical properties of the model \cite{Swendsen87, Wolff89}. Numerically, as proposed by Coniglio and Klein, geometrical clusters can be obtained starting with magnetic clusters, defined as sets of neighboring spins in the same state \cite{Fisher67, Fisher67b, Stoll72}, and diluting bonds in such a way that, at criticality, their percolation-like properties are equivalent to the magnetic ones \cite{Sykes76, Coniglio80, Coniglio82c}.

The geometrical properties of Coniglio-Klein clusters have been intensively studied obtaining several numerical and exact results \cite{Coniglio89, Asikainen03}. However, the transport properties for $q>1$ are, to the best of our knowledge, still unexplored. In this work, we study the conductivity of these clusters for values of $q$ between zero and four in two dimensions. Given the self-similar properties of the clusters at criticality for $q\leq4$, we assume that the conductivity $C$ between two points scales as,
\begin{equation}
\label{eq::conductivity_power_law}
C(r) \sim r^{-\frac{t}{\nu}},
\end{equation}
where $r$ is the distance between the points, $t$ the conductivity exponent, and $\nu$ the critical exponent of the correlation length. $t/\nu$ is then the reduced conductivity exponent, the focus of our work.

For bond percolation ($q=1$), the conductivity has been studied with different methods as, e.g., diffusion methods \cite{Hong84}, transfer matrix methods \cite{Derrida82, Zabolitzky84, Normand88}, and star-triangle transformations \cite{Lobb84, Frank88, Grassberger99}. Some conjectures which have been proposed for the conductivity exponent \cite{Alexander82, Aharony84} were ruled out by numerical data \cite{Normand88}. In our study, we extend the calculation of the conductivity exponent to $q=1.5,2,2.5,3,3.5$, and $4$. 

The paper is organized as follows. In Section \ref{section::Model} we describe the $q$-state Potts model and its relation with bond percolation. In Section \ref{section::Simulations} we present the two procedures to estimate the value of the reduced conductivity exponent. In Section \ref{section::Results} we report our results which are then discussed in Section \ref{section::Discussion}. The concluding remarks are in Section \ref{section::Conclusion}.

\section{\label{section::Model}Model}

In the $q$-state Potts model on a graph, each node is a spin and can assume $q$ different states, $\sigma=0,1,..., q-1$. The Hamiltonian of the model is,
\begin{equation}
\label{eq::Potts_Hamiltonian}
\mathcal{H}=-\sum_{\langle i, j\rangle} J_{ij} \left( \delta_{\sigma_i \sigma_j} -1 \right),
\end{equation}
where $\delta_{\sigma_i \sigma_j}$ is unity if $\sigma_i=\sigma_j$ and zero otherwise, $J_{ij}$ is the coupling strength between spins $i$ and $j$, and the summation is over all pairs of interacting spins. For simplicity, we assume only interactions among nearest neighbors with the same coupling strength, $J_{ij} \equiv J$.

A geometrical description of the model which reproduces the scaling behavior of the magnetic transition is provided by the Kastelyn-Fortuin theorem \cite{Fortuin72}. The partition function associated to the $q$-state Potts model can be written as a sum over all possible bond configurations  $\{ \nu \}$,
\begin{equation}
\label{eq::KF_clusters_partition_function}
Z=\sum_{\lbrace \nu \rbrace}  q^{N_c} p^b (1-p)^{\left(N_b-b \right)},
\end{equation}
where $N_b$ is the total number of bonds, and $b$ and $N_c$ are, respectively, the number of occupied bonds and clusters in the configuration $\nu$. This is equivalent to a sum over configurations of bond percolation in the original graph, weighted by a factor of $q^{N_c}$, and, consequently, for $q=1$, one recovers the generating function of the random bond percolation model. Bonds are established between neighboring spins in the same state with probability $p=1-e^{-K}$, where $K=\frac{J}{k_B T}$, $T$ is the temperature, and $k_B$ is the Boltzmann constant. Coniglio-Klein clusters are then defined as sets of spins (nodes) connected through these bonds. This equivalence has been useful in the development of efficient Monte-Carlo procedures where the critical slowing down is significantly reduced \cite{Swendsen87, Wolff89}.

\begin{figure}[h]
\begin{center}
\includegraphics[width=0.8\columnwidth]{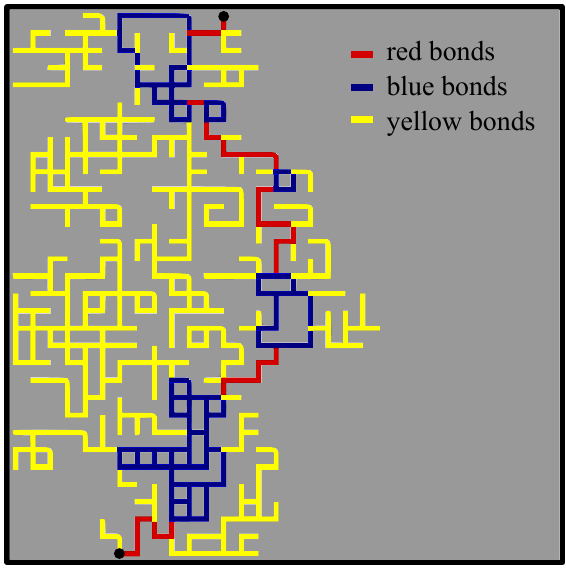}
\vspace*{-7mm}
\end{center}
\caption{
\label{fig::backbone_red_bonds}
(color online) Largest cluster of a configuration for $q=1$ and a lattice size of $32$. The backbone between the two dotted vertices is formed by blobs of bonds belonging to parallel paths (blue (dark) bonds) linked together by red (grey) bonds. The other bonds (yellow (white) bonds) are dangling ends.
\vspace*{-5mm}
}
\end{figure}

Let us consider now any two points on the largest cluster. One can define the backbone as the set of all possible paths joining these two points. All the bonds of the largest cluster outside the backbone are dangling ends. In the backbone, one can distinguish the red from the blue bonds: red bonds are singly connected bonds that, if one of them is removed, splits the cluster into two, while all the others are blue bonds, see Fig. \ref{fig::backbone_red_bonds}. The typical picture of the largest cluster at criticality is a set of blue bonds forming blobs, i.e., sets of parallel paths, linked together by red bonds \cite{Coniglio81, Pike81, Herrmann84b}.

Describing the bonds as resistors, one can analyze the transport properties of the cluster. In this work, we apply a potential difference between two points of the largest cluster, separated by a distance $r$, and we solve the Kirchhoff's law for the network of resistors. We characterize the dependence of the conductivity on the distance between the points and compute the conductivity exponent for different values of $q$.

\section{\label{section::Simulations}Simulations}

We simulate the $q$-state Potts model on a square lattice with periodic boundary conditions in vertical and horizontal directions. The Coniglio-Klein clusters are generated using the Swendsen-Wang algorithm \cite{Swendsen87}. As described above, critical clusters are identified by establishing bonds between neighboring spins in the same state, with probability $p_c=1-e^{-K_c}$, where $K_c=\frac{J}{k_B T_c}$ and $T_c$ is the critical temperature \cite{Hoshen76}. We focus on the largest cluster without periodic boundary conditions. Therefore we compute the part of the largest cluster that is in the middle of the lattice. In order to do so, we look for a node belonging to the largest cluster, starting from the nearest point that is down right from the center of the system. If it does not belong to the largest cluster, we continue, going through the nodes of the lattice by scanning the lattice in two directions at the same time: starting from the middle point, we go up from left to right and down from right to left. When we reach a node that belongs to the largest cluster, we clip the fraction of the largest cluster, without periodic boundary conditions, containing this node. We then identify the lowest and the highest node: the lowest node is defined as the first node belonging to the cluster when one scans through the lattice from the bottom to the top (left to right), whereas the highest node is the first node belonging to the cluster when scanning through from the top to the bottom (right to left). Since the current vanishes in the dangling ends and only flows through the backbone, we restrict our calculation to the latter, which can be identified with the burning method proposed by Herrmann \textit{et al.} \cite{Herrmann84}. Using Kirchhoff's laws, for every node $i$ in the backbone, we have
\begin{equation}
\label{eq::Kirchhoff_law}
\sum_{j} C_{ij} \left( V_i - V_j \right) =0,
\end{equation}
where the sum is over all neighboring spins $j$, and the conductivity $C_{ij}$ between nodes $i$ and $j$, is unity if there is a bond between them or zero otherwise. We impose the boundary condition $V=L^2$ for the highest node and zero potential for the lowest one. We then invert the (sparse) conductivity matrix, with elements $C_{ij}$, using a sparse matrix solver \footnote{To compute the solution of the sparse matrix equation, we use the Intel MKL Direct Sparse Solver} and obtain the potential for every node as well as the global conductivity. We compute the dependence of the conductivity on the distance $r$. For $q <4$, we show that the conductivity $C$ as a function of the distance $r$ is a power law and we estimate the reduced conductivity exponent $D_{\sigma}$, given by 
\begin{equation}
\label{eq::conductivity_power_law_direct_measurement}
C(r) \sim r^{-D_{\sigma}}.
\end{equation}
The value $V=L^2$ has been chosen to avoid the sparse matrix solver to deal with small potentials, as the backbone mass is scaling with an exponent larger than one \cite{Deng04}.

Simulations were performed on lattices of size \mbox{$L=32,64,128,256,512,1024,2048$}, for $q$ between one and three, and for $q=1$ a lattice size of $L=4096$ was also studied. For $q=4$, only system sizes up to $1024$ where considered. The number of samples generated for each lattice size ranges from $10^8$ for the smallest system sizes till about $10^4$ for the largest system sizes. As the analyses of the conductivity and random walks are computationally demanding, we discarded samples between measurements to remove correlated samples that would not improve the precision of the results. To decorrelate from the initial configuration, we reject the first $2L$, $3L$, and $16L$ sweeps, for $q=2$, $3$, and $4$, respectively. Between measurements, we reject eight, $16$, and $64$ sweeps, for $q=2$, $3$, and $4$, respectively. We then use a binning procedure and the integrated autocorrelation time from Ref. \cite{Deng07} to obtain the error bars. For $q=1$, since all spins are in the same state, each generated configuration is a random bond percolation configuration independent from the previous one.

\begin{figure*}
       \centering
       \begin{tabular}{lll}
               \includegraphics[width=0.67\columnwidth]{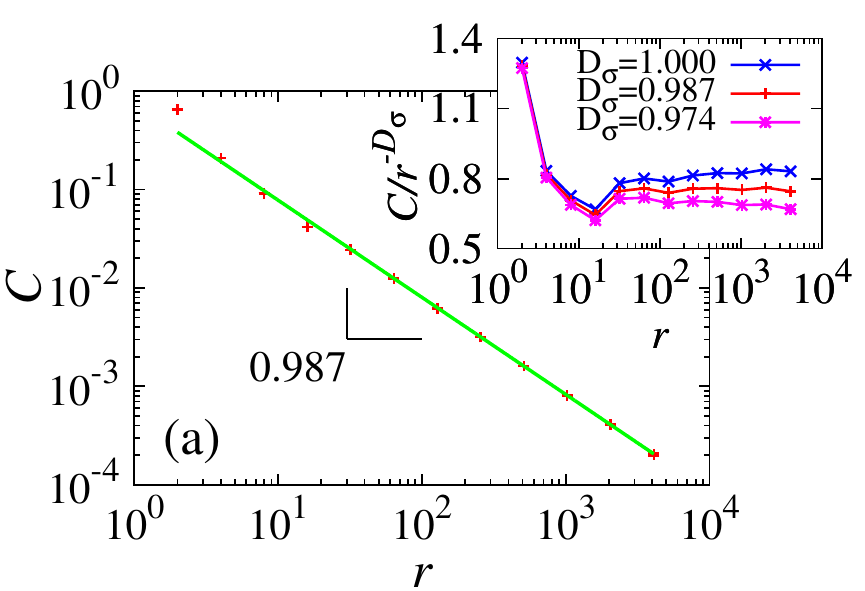}&
               \includegraphics[width=0.67\columnwidth]{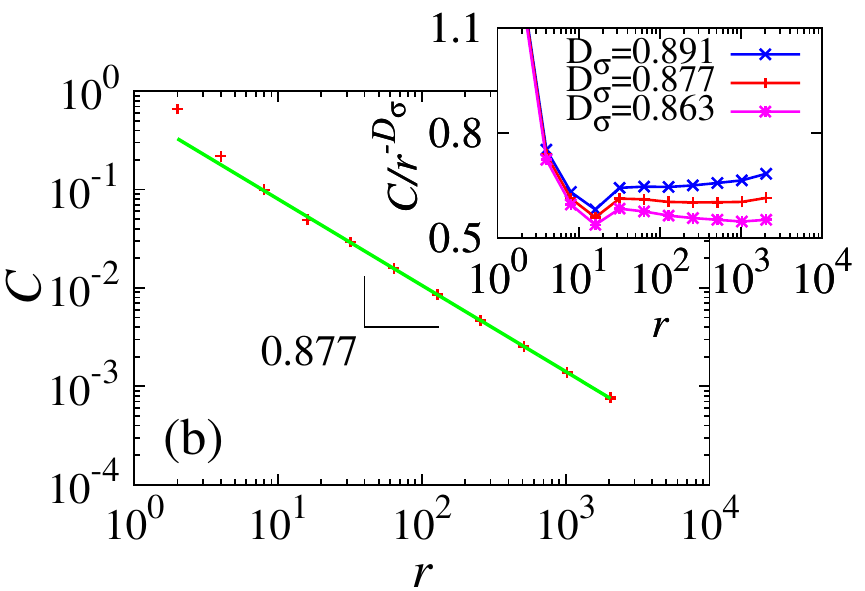}&
               \includegraphics[width=0.67\columnwidth]{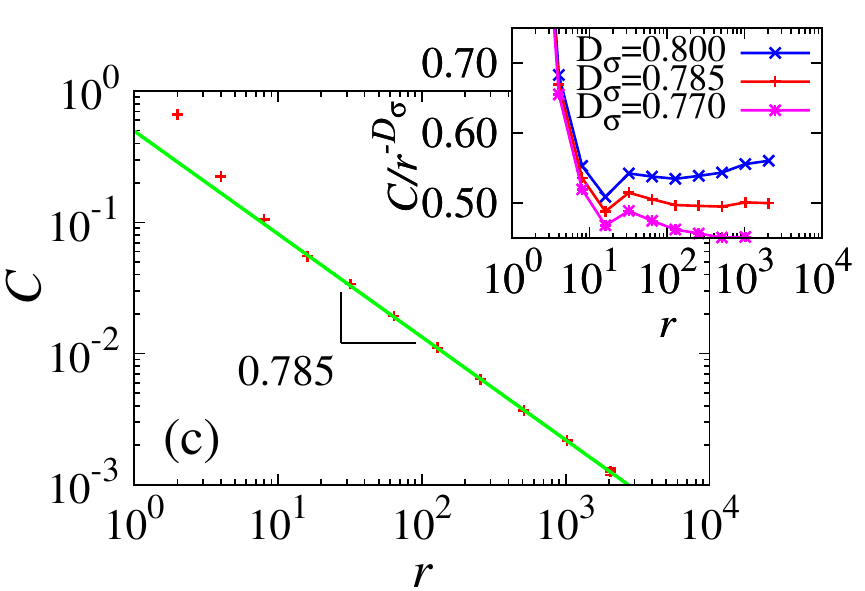}\\
      \end{tabular}
      \vspace*{-4mm}
       \caption{
       \label{fig::figure_conductivity_exponent_q_1_2_3}      
      (color online) Conductivity $C$ as a function of the distance $r$ between the highest and lowest points of the largest cluster for (a) $q=1$, (b) $q=2$, and (c) $q=3$. The solid lines are guides to the eye of the form $C(r)=b_C r^{-D_{\sigma}}$. In the inset, the rescaled conductivity $C(r)/r^{-D_{\sigma}}$ is plotted as a function of the distance $r$ for different values of the reduced conductivity $D_{\sigma}$.
       }
       \vspace*{-2mm}
\end{figure*}
\begin{figure*}
       \centering
       \begin{tabular}{lll}
               \includegraphics[width=0.67\columnwidth]{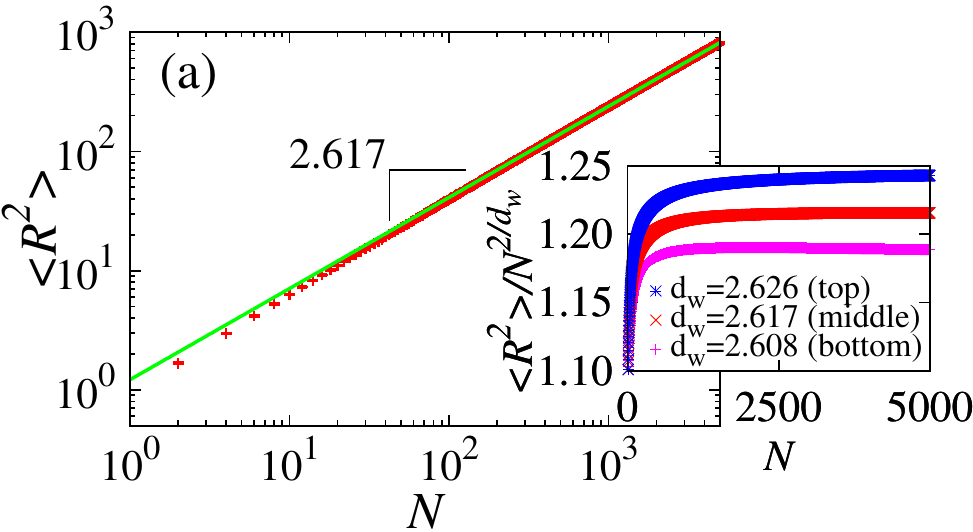}&
               \includegraphics[width=0.67\columnwidth]{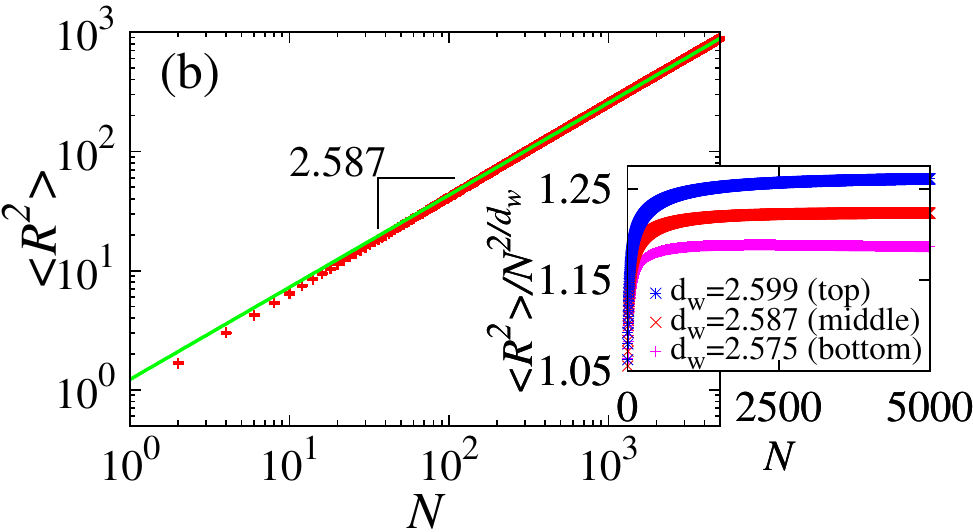}&
               \includegraphics[width=0.67\columnwidth]{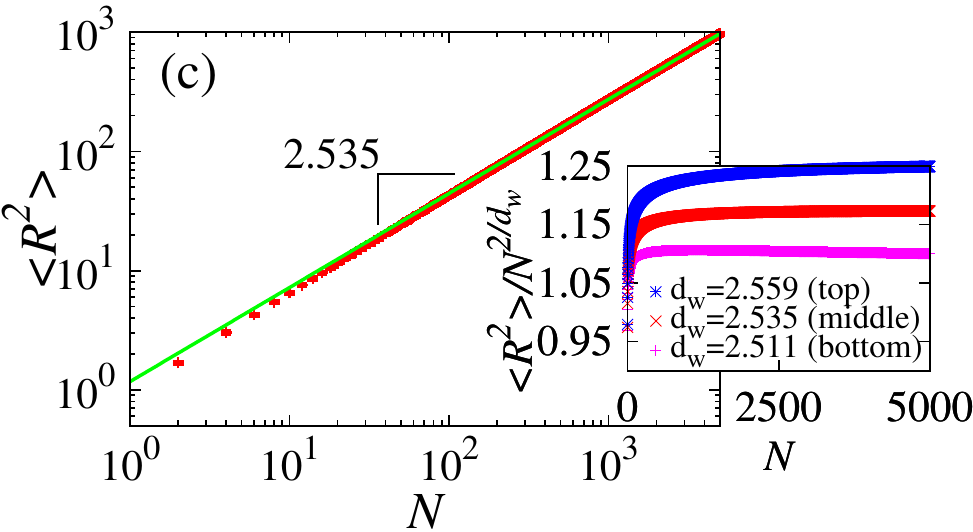}\\
       \end{tabular}
       \vspace*{-3mm}
       \caption{
       \label{fig::figure_mean_square_displacement_q_1_2_3}
(color online) Mean square displacement of the random walk $\langle R^2 \rangle$ as a function of the number of steps $N$ for (a) $q=1$, (b) $q=2$, and (c) $q=3$. The solid lines are guides to the eye of the form $\langle R^2 (N)\rangle=b_R N^{\frac{2}{d_w}}$. In the inset, the rescaled mean square displacement $\langle R^2(N) \rangle / N^{\frac{2}{d_w}}$ is plotted as a function of the number of steps $N$ for different values of the random walk exponent $d_w$.
\vspace*{-4mm}
       }
\end{figure*}

We also implemented a procedure based on the diffusion of a random walk on the backbone to estimate the reduced conductivity exponent. Such diffusion can be related with the transport properties since, on a random fractal medium, the walker's mean square displacement $\langle R^2(N) \rangle$ relates with the number of steps $N$ as 
\begin{equation}
\label{eq::mean_square_radius_of_the_walker}
\langle R^2(N) \rangle \sim N^{\frac{2}{d_w}},
\end{equation}
where $d_w$ is the random walk dimensionality. The mean probability to return to the initial position scales as 
\begin{equation}
\label{eq::mean_prob_of_return_of_the_walker}
\langle P_0(N) \rangle \sim N^{-\frac{d_f}{d_w}} \sim N^{-\frac{d_s}{2}},
\end{equation} 
where $d_f$ is the fractal dimension of the backbone and $d_s$ is the fracton dimension \cite{Alexander82}. The reduced conductivity exponent can then be computed from the identity \cite{Stanley84},
\begin{equation}
\label{eq::conductivity_RW_backbone}
\left(\frac{t}{\nu}\right)_{\mathrm{rw}}=d_w-d_f=d_w\left(1-\frac{d_s}{2} \right).
\end{equation}
The diffusion of a random walk on the backbone can be computed exactly following the algorithm proposed in Ref. \cite{Hong84}. Once the largest cluster and the lowest and highest points have been identified with the method described before, one chooses an initial point in the middle of the backbone and restricts the diffusion process to a constant chemical length from this point. To do so, one takes the site of the largest cluster which is closer to the middle of the line joining the lowest and the highest point. The diffusion is solely considered on the subset of sites in the largest cluster which are distant from the middle point less than a maximum chemical length. This distance is defined as the minimum number of steps for the walker to go from one site to the other. At each step, for every node in the backbone, one computes exactly the probability of the walker to be at this node and we measure $\langle R^2(N) \rangle$ and $\langle P_0(N) \rangle$. Following this procedure, we performed simulations on lattices of lateral size $L=1024$, averaging results over $10^4$ configurations. We limited the walker to a maximum chemical distance of $300$ steps from the origin of the walk. With $d_w$ and $d_s$, we deduced the value of the reduced conductivity exponent using Eq.~(\ref{eq::conductivity_RW_backbone}).

\section{\label{section::Results}Results}

We start discussing the cases \mbox{$q=\{1,2,3\}$}, followed by the special values $q=4$ and $q=0$. For all cases, we compare the estimates for the reduced conductivity obtained from both the conductivity measurement $D_{\sigma}$ and the random walk method $(t/\nu)_{\mathrm{rw}}$. We use the results from both methods to estimate the reduced conductivity $t/ \nu$ for different values of $q$. We also present results on the reduced conductivity exponent for non-integer values of $q$, namely, $q=1.5,2.5,3.5$.

\subsection{\label{q=1,2,3}$q=1,2,3$}

\begin{table}
\vspace*{-4mm}
\caption{\label{tab::results_q_1_2_3}Numerical results for the reduced conductivity exponent $D_\sigma$, random walk dimension $d_w$ and fracton dimension $d_s$. $(t / \nu)_{\mathrm{rw}}$ is the reduced conductivity exponent obtained with Eq.~(\ref{eq::conductivity_RW_backbone}).}
\begin{center}
\vspace*{-4mm}
\begin{tabular}{c c c c c c}   
\hline \hline
   $q$ & $\nu$ & $D_{\sigma}$ & $d_w$ & $d_s$ & $\left(\frac{t}{\nu}\right)_{\mathrm{rw}}$ \\
   \hline
   
   $1$ & $\frac{4}{3}$ & $0.987 \pm 0.013$ &  $2.617 \pm 0.009$ &  $1.247 \pm 0.007$ & $0.985 \pm 0.013$  \\
   $2$ & $1$  & $0.877 \pm 0.014$ &  $2.587 \pm 0.012$ & $1.325 \pm 0.014$ & $0.873 \pm 0.022$  \\
   $3$ & $\frac{5}{6}$ & $0.785 \pm 0.015$ &  $2.535 \pm 0.024$ & $1.396 \pm 0.025$ & $0.766 \pm 0.039$ \\
   
   \hline
   \hline

\end{tabular}
\vspace*{-7.5mm}
\end{center}
\end{table}

\begin{figure*}
       \centering
       \begin{tabular}{lll}
               \includegraphics[width=0.67\columnwidth]{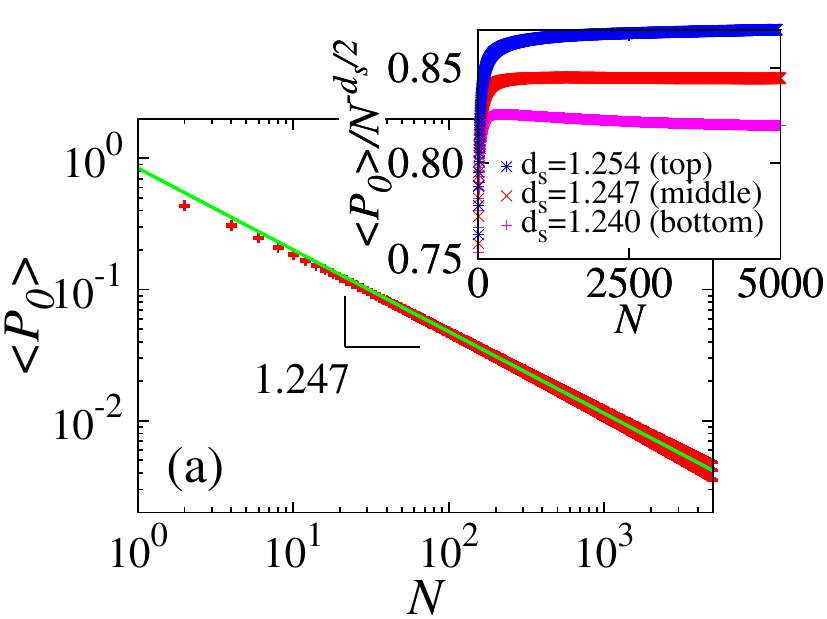}&
               \includegraphics[width=0.67\columnwidth]{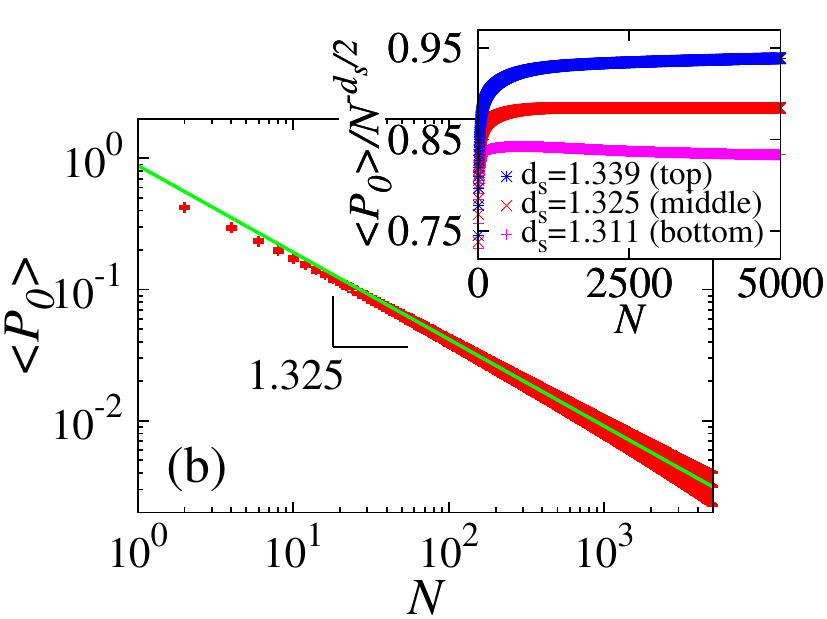}&
               \includegraphics[width=0.67\columnwidth]{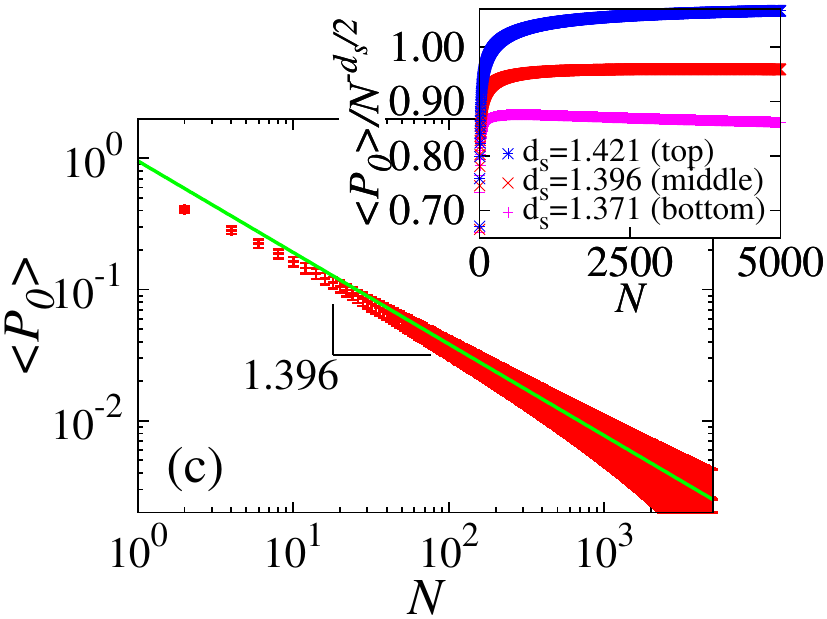}\\
       \end{tabular}
       \vspace*{-4mm}
       \caption{
       \label{fig::figure_return_probability_q_1_2_3}
      (color online) Mean probability to return to the initial position $\langle P_0 \rangle$ as a function of the number of steps $N$ for (a) $q=1$, (b) $q=2$, and (c) $q=3$. The solid lines are guides to the eye of the form $\langle P_0 \rangle =b_P N^{-\frac{d_s}{2}}$. In the inset, the rescaled return probability $\langle P_0 (N)\rangle / N^{-\frac{d_s}{2}}$ is plotted as a function of the number of steps $N$ for different values of the fracton dimension $d_s$. 
      \vspace*{-5mm} 
       }
\end{figure*}

Figure \ref{fig::figure_conductivity_exponent_q_1_2_3} shows the dependence on the distance $r$ of the conductivity obtained with the sparse matrix solver method. In the insets, we show the rescaled conductivity, $C(r)/r^{-D_{\sigma}}$ for different values of $D_{\sigma}$. To reduce the influence of finite-size effects, the value of the reduced conductivity exponent $D_{\sigma}$ and respective error bar are estimated by checking for which interval of values for $D_{\sigma}$ a plateau is asymptotically observed for $C(L)/L^{-D_{\sigma}}$. We obtained the values of $D_{\sigma}$ for $q=1$, $2$, and $3$, as summarized in Table \ref{tab::results_q_1_2_3}, where error bars are estimated from the asymptotic behavior in the insets of Fig. \ref{fig::figure_conductivity_exponent_q_1_2_3}. Figures \ref{fig::figure_mean_square_displacement_q_1_2_3} and \ref{fig::figure_return_probability_q_1_2_3} show, respectively, the data for the mean square displacement $\langle R^2(N) \rangle$ and the probability of return $\langle P_0(N) \rangle$. Given the proposed relations Eqs.~(\ref{eq::mean_square_radius_of_the_walker})~and~(\ref{eq::mean_prob_of_return_of_the_walker}), we obtain $d_w$ and $d_s$. 
From Eq.~(\ref{eq::conductivity_RW_backbone}), we estimate the reduced conductivity exponent $(t/ \nu)_{\mathrm{rw}}$. All results are summarized in Table \ref{tab::results_q_1_2_3}.

\begin{figure}[h]
\begin{center}
\includegraphics[width=0.8\columnwidth]{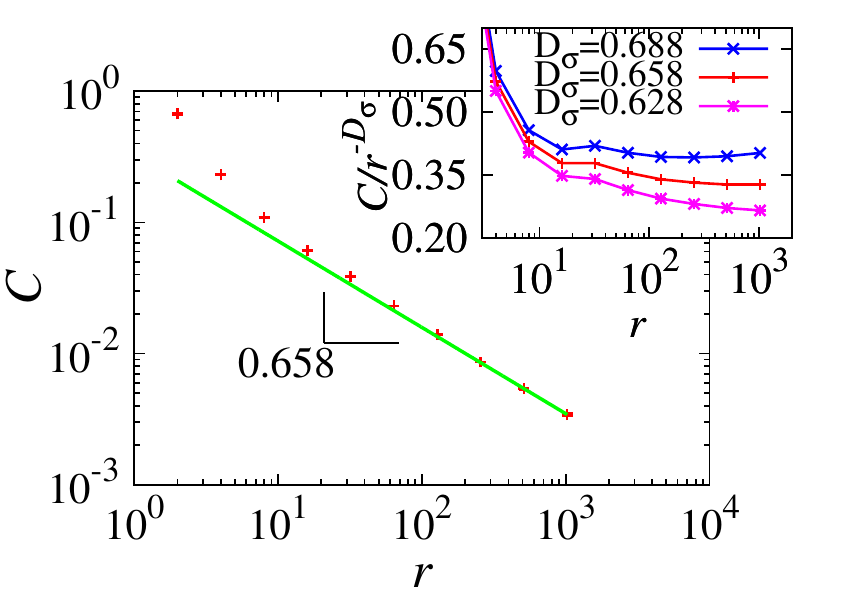}
\vspace*{-8mm}
\end{center}
\caption{
\label{fig::Potts_q_4_conductivity}
(color online) Conductivity $C$ as a function of the distance $r$, for $q=4$. The line corresponds to the fit with the power-law \textit{Ansatz}, in Eq.~(\ref{eq::conductivity_power_law_direct_measurement}), where \mbox{$D_{\sigma}=0.658$}. In the inset, the rescaled conductivity $C(r)/r^{-D_{\sigma}}$ is plotted as a function of the distance $r$ between the highest and the lowest points of the largest cluster for different values of $D_{\sigma}$.
\vspace*{-5mm}
}
\end{figure}

The results obtained with the sparse matrix solver and with the random walk method are consistent with each other. Overlapping the confidence intervals obtained with both methods, we estimate the value of the reduced conductivity to be $t/\nu=0.986 \pm 0.012$, $t/\nu=0.877 \pm 0.014$, and $t/\nu=0.785 \pm 0.015$, for $q=1$, $2$, and $3$, respectively. 
For $q=1$, random bond percolation is recovered. Our result in this limit is consistent with the ones previously reported in the literature using different methods $0.9745 \pm 0.0015$ \cite{Normand88}, $0.977 \pm 0.010$ \cite{Frank88}, $0.9826\pm 0.0008$\cite{Grassberger99}.

\subsection{\label{q=4}$q=4$}

In two dimensions, the nature of the magnetic transition in the $q$-state Potts model crosses over from second order, for $q \leq 4$, to first order, for $q>4$, and, therefore, logarithmic prefactors are typically observed for $q=4$. For example, several works based on renormalization group theory show that such prefactors are necessary for the specific heat, spontaneous magnetization, and susceptibility \cite{Cardy80, Salas97}. 
\begin{figure}[h]
\begin{center}
\includegraphics[width=0.8\columnwidth]{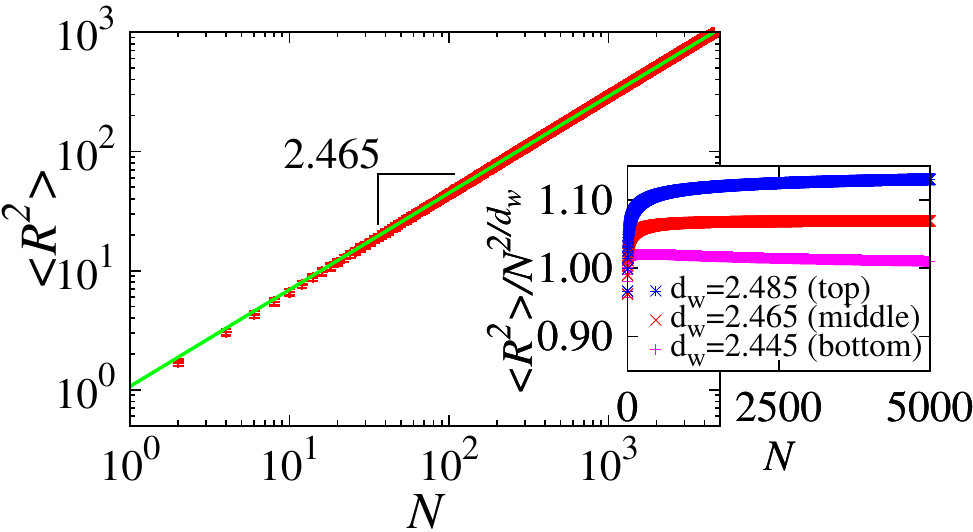}
\vspace*{-8mm}
\end{center}
\caption{
\label{fig::Potts_q_4_RW_exponent_dw}
(color online) Mean square displacement of the random walk $\langle R^2 (N) \rangle$ as a function of the number of steps $N$, for $q=4$. The solid line is a guide to the eye of the form \mbox{$\langle R^2 (N) \rangle = b_R N^{\frac{2}{d_w}}$} where $d_w=2.465$. In the inset, the rescaled mean square displacement $\langle R^2 (N) \rangle / N^{\frac{2}{d_w}}$ is plotted as a function of the number of steps $N$ for different values of $d_w$.
\vspace*{-2mm}
}
\end{figure}
We compared, for $q=4$, the power-law scaling of Eq.~(\ref{eq::conductivity_power_law_direct_measurement}) with a power-law including logarithmic prefactors of the form \mbox{$C(r) \sim r^{-D_{\sigma}} |\log(r)|^{\tilde{D_{\sigma}}}$}. We did not observe an improvement of the fit including the logarithmic prefactor. Therefore, we used the same power-law behavior as considered for $q<4$.

\begin{figure}[h]
\begin{center}
\vspace*{-2mm}
\includegraphics[width=0.75\columnwidth]{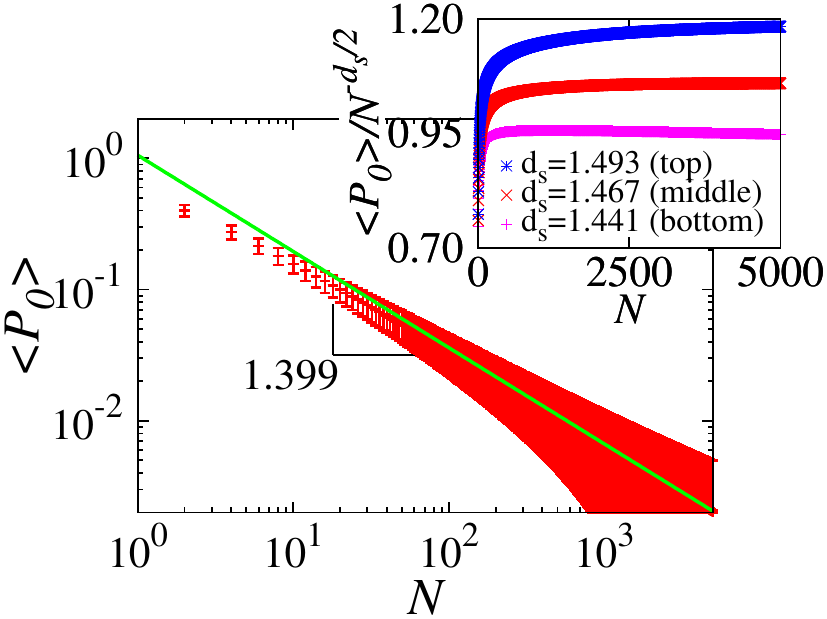}
\vspace*{-8mm}
\end{center}
\caption{
\label{fig::Potts_q_4_RW_exponent_ds}
(color online) Mean probability to return to the initial position $ \langle P_0 \rangle $ as a function of the number of steps $N$, for $q=4$. The solid line is a guide to the eye of the form $ \langle P_0 (N) \rangle = b_P N^{\frac{-d_s}{2}}$ where $d_s=1.467$. In the inset, the rescaled probability of return $\langle P_0 (N) \rangle / N^{\frac{-d_s}{2}}$ is plotted as a function of the number of steps $N$ for different values of $d_s$.
\vspace*{-3mm}
}
\end{figure}

Figure \ref{fig::Potts_q_4_conductivity} shows the conductivity as a function of $r$, giving $D_{\sigma}= 0.658 \pm 0.030$. From Figs. \ref{fig::Potts_q_4_RW_exponent_dw} and \ref{fig::Potts_q_4_RW_exponent_ds} we obtain $d_w=2.465 \pm 0.020$ and $d_s=1.467 \pm 0.026$ and using Eq.~(\ref{eq::conductivity_RW_backbone}) \mbox{$\left(t / \nu \right)_{\mathrm{rw}}=0.657 \pm 0.037$}. The results obtained with the sparse matrix solver and with the random walk are in agreement. From both methods we estimate \mbox{$t/ \nu = 0.658 \pm 0.030$}.

\subsection{\label{q=0}$q=0$}

As $q \rightarrow 0$ the critical temperature $T_c$ diverges and, consequently, the critical probability $p_c$ vanishes. As a result, no criticality is observed at finite temperature for $q=0$ \cite{Wu78b}. However, it is possible to show that in some limits,
the partition function for $q=0$ is equivalent to the one of the uniform spanning trees \cite{Jacobsen05}, i.e., a sum over the ensemble of all possible spanning trees. Here we study the conductivity of spanning trees. 

\begin{figure}[h]
\begin{center}
\vspace*{-0mm}
\includegraphics[width=0.8\columnwidth]{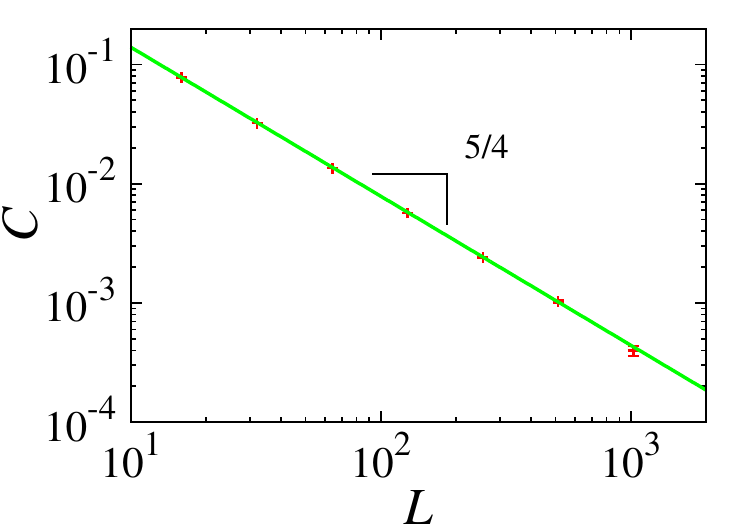}
\end{center}
\vspace*{-7mm}
\caption{\label{fig::Potts_q_0_conductivity_exponent} (color online) Conductivity $C$ as a function of the lattice size $L$, for $q=0$. The solid line is a guide to the eye of the form $C(L)=b_C L^{-D_{\sigma}}$, where $D_{\sigma}=5/4$.}
\vspace*{-4mm}
\end{figure}

The backbone between any two points on a spanning tree is a single fractal path of fractal dimension $5/4$ \cite{Deng04}, an exponent that can be obtained exactly, for example, from the size dependence of the number of red bonds \cite{Coniglio89, Scholder09}. If each bond is a resistor with the same resistance, we expect the fractal dimension of the conductivity to be the same as the one of the shortest path (and the backbone), i.e., $D_{\sigma}=5/4$. In Figure \ref{fig::Potts_q_0_conductivity_exponent} we show that this result still holds when we consider a spanning tree where all the points in the top row ($V=L^2$) are connected with all the points in the bottom one ($V=0$). 

\subsection{\label{q=1.5,2.5,3.5}$q=1.5,2.5,3.5$}

\begin{figure*}
       \centering
       \begin{tabular}{lll}
               \includegraphics[width=0.67\columnwidth]{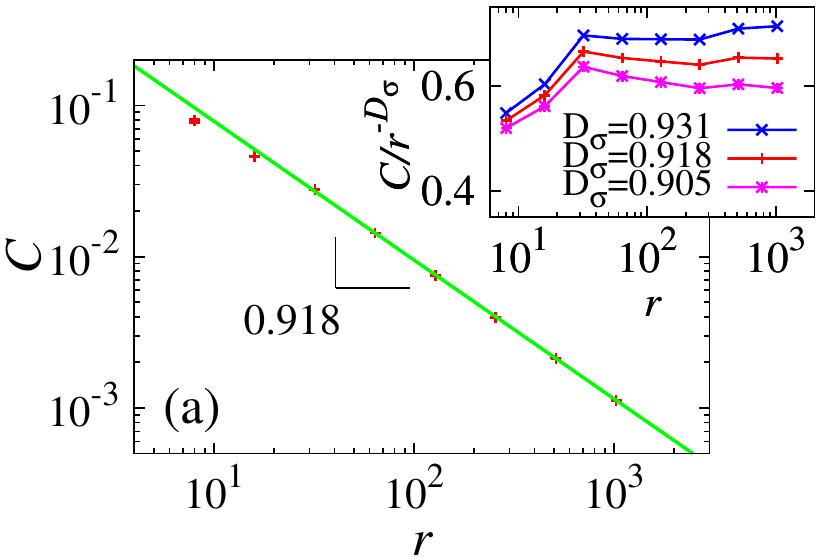}&
               \includegraphics[width=0.67\columnwidth]{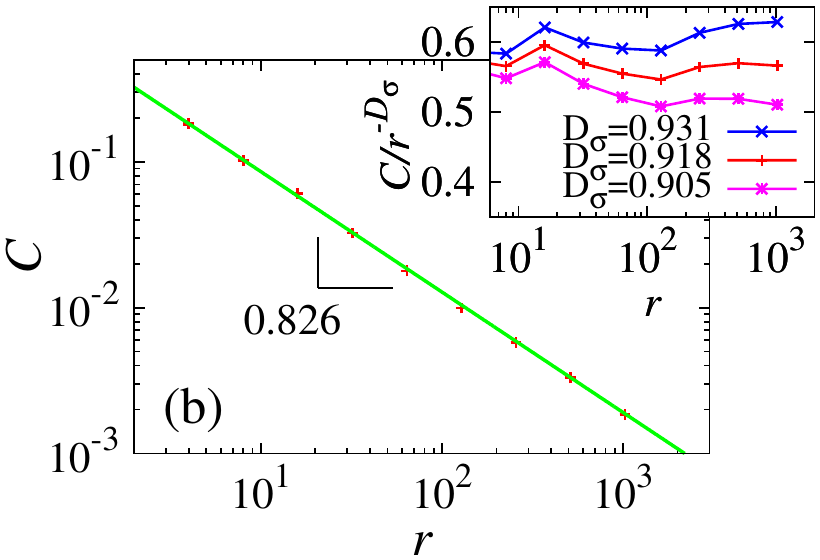}&
               \includegraphics[width=0.67\columnwidth]{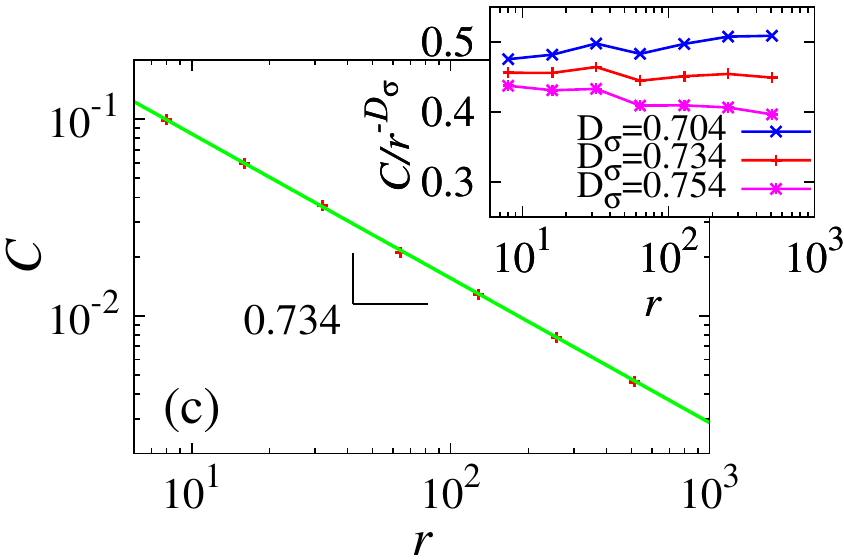}\\
      \end{tabular}
      \vspace*{-4mm}
      \caption{
     \label{fig::figure_conductivity_exponent_q_1p5_2p5_3p5}      
     (color online) Conductivity $C$ as a function of the distance $r$ between the highest and lowest points of the largest cluster for (a) $q=1.5$, (b) $q=2.5$, and (c) $q=3.5$. The solid lines are guides to the eye of the form $C(r)=b_C r^{-D_{\sigma}}$. In the inset, the rescaled conductivity $C(r)/r^{-D_{\sigma}}$ is plotted as a function of the distance $r$ for different values of the reduced conductivity $D_{\sigma}$.
       }
       \vspace*{-2mm}
\end{figure*}

\begin{figure}[h]
\begin{center}
\includegraphics[width=0.8\columnwidth]{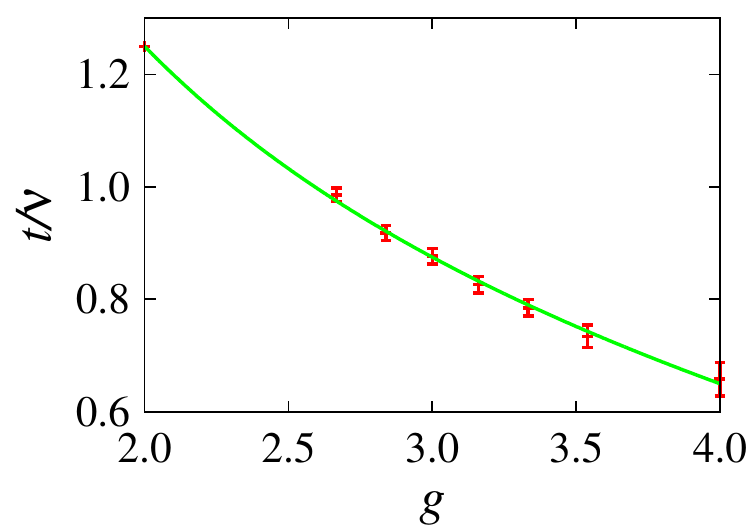}
\end{center}
\vspace*{-7mm}
\caption{ \label{fig::Potts_conductivity_exponent_Ansatz_t} (color online) Reduced conductivity exponent $t / \nu$ as a function of the Coulomb gas parameter $g$ (data points), and conjecture (solid line).}
\end{figure}

To obtain intermediary values of the reduced conductivity exponent for non-integer values of $q$ we have computed the conductivity of Coniglio-Klein clusters for $q=1.5$, $2.5$, and $3.5$. We generated the bond configuration using the Chayes-Machta algorithm \cite{Chayes98}, which is a generalization of the Swendson-Wang algorithm. The simulations were done for $q=1.5$ and $2.5$ for lattice sizes between $16$ and $1024$, and for $q=3.5$ between $16$ and $512$. Results are averages over $10^5$ samples for the smallest system sizes to $10^3$ for the largest ones. For $q=2.5$, and $3.5$, respectively, we rejected the first $3L$ and $16L$ sweeps to thermalize and $26 $ and $358$ configurations between two consecutive measurements \cite{Deng07}. The estimates of the reduced conductivity exponent obtained in Fig. \ref{fig::figure_conductivity_exponent_q_1p5_2p5_3p5} are $D_{\sigma}=0.918 \pm 0.013$, $D_{\sigma}=0.826 \pm 0.015$ and $D_{\sigma}=0.734 \pm 0.020$ for $q=1.5$, $q=2.5$, and $q=3.5$, respectively.

\section{\label{section::Discussion}Discussion}

In Table \ref{tab::results}, we summarize the results for the conductivity exponents for $q\leq 4$. For $q>4$, the magnetic transition is first order in nature and the backbone (as the largest cluster) is compact \cite{Deng04}. Therefore, asymptotically, no dependence of the conductivity on the distance between two points is expected. We have confirmed this behavior with simulations for $q=5$ (not shown). 

Our estimates of the reduced conductivity are monotonically decreasing with $q$. This is in line with the results for the dimension of the backbone, which increases with $q$ and approaches the spatial dimension for $q=4$ \cite{Deng04}. As the dimension of the backbone increases, it becomes a dense object and the conductivity exponent vanishes, being zero for $q>4$. 

The conductivity $C$ as a function of the distance $r$ shows a bump in the reduced conductivity $C(r) / r^{-D_{\sigma}}$ around $r=16$ for $q$ between one and four. This behavior was also previously observed for $q=1$ on the square lattice, with the star-triangle transformation, and is not due to statistical fluctuations \cite{Lobb84, Frank88, Grassberger99}.

We have only considered integer values of $q$ in the range between zero and four. However, the Potts model can be defined for any real value of $q$ \cite{Fortuin72}. From the obtained conductivity exponents we can conjecture a dependence on the value of $q$. The relation between the $q$-state Potts model and the Coulomb gas theory \cite{Nienhuis84, diFrancesco87} has been very useful to develop exact relations for the red bond and hull fractal dimensions \cite{Saleur87, Coniglio89}, and also to conjecture the shortest path fractal dimension \cite{Deng10}. The Coulomb gas coupling $g$ is related with $q$ by, 

\begin{equation}
\label{eq::g_coupling_constant}
q=2+2 \cos\left(\frac{g\pi}{2} \right),
\end{equation}
with $2 \leq g \leq 4$. From the obtained results we propose the following conjecture for the dependence of the conductivity exponent on $g$, 
\begin{equation}
\label{eq::conjecture_reduced_conductivity_exponent}
\frac{t}{\nu}(g)=\frac{9}{5g}+\frac{1}{2}-\frac{3}{40}g.
\end{equation}
This conjecture fits, within the error bars, the data points in Fig. \ref{fig::Potts_conductivity_exponent_Ansatz_t} and displays a continuous behavior in the limit $q \rightarrow 0$, where it converges to the value $5/4$. It was obtained by a weighted least-square fit, under the assumption that the reduced conductivity exponent takes the form $\frac{t}{\nu}(g)=ag+b+c/g$, with $a$, $b$, and $c$ rational, and that it converges towards $5/4$ for $q \rightarrow 0$. The values of the constants $a$, $b$, and $c$ have been obtained by fitting the equation to our results.

\begin{table}
\vspace*{-3mm}
\caption{\label{tab::results}Values of the conjectured conductivity exponent $t / \nu (g)$ and the measured reduced conductivity exponent $t / \nu$ for $q=0,1,1.5,2,2.5,3,3.5,4$ and the corresponding Coulomb gas parameter $g$.}
\begin{center}
\vspace*{-2mm}
\begin{tabular}{c c c c}   \hline \hline
   $q$ & $g$ & $\frac{t}{\nu}(g)$ & $\frac{t}{\nu}$ \\
   \hline
   $0$ & $2$ & $\frac{5}{4}$ & $\frac{5}{4}$  \\
   $1$ & $\frac{8}{3}$ & $0.975$ & $0.986 \pm 0.012$ \\
   $1.5$ & $2.8391$ & $0.921$ & $0.918 \pm 0.013$ \\
   $2$ & $3$ & $0.875$ & $0.877 \pm 0.014$  \\
   $2.5$ & $3.1609$ & $0.832$ & $0.826 \pm 0.015$ \\
   $3$ & $\frac{10}{3}$ & $0.790$ & $0.785 \pm 0.015$ \\
   $3.5$ & $3.5399$ & $0.743$ & $0.734 \pm 0.020$ \\
   $4$ & $4$ & $0.650$ & $0.658 \pm 0.030$ \\
   \hline
   \hline
\end{tabular}
\end{center}
\vspace*{-4mm}
\end{table}

\section{\label{section::Conclusion}Conclusion}

We implemented two independent methods to compute the reduced conductivity exponent $t / \nu$ for different values of $q$ in the $q$-state Potts model. In the first method we compute directly the conductivity on Coniglio-Klein clusters whereas in the second one we compute exactly the diffusion of a random walk on the backbone of these clusters. The results obtained with both methods are in agreement with each other. From the data of the values of the reduced conductivity exponent $t / \nu$ for integer values of $q$, and for the intermediary values $q=1.5,2.5$ and $3.5$, we propose a conjecture which interpolates from our results the conductivity for any real value of $q$ between zero and four.

\begin{acknowledgments}
We acknowledge financial support from the ETH Risk Center and the Brazilian Institute INCT-SC.
\end{acknowledgments}

\bibliography{bibliography}

\end{document}